\documentclass[twocolumn,prb,showpacs,preprintnumbers]{revtex4}

\usepackage{graphicx}
\usepackage{hyperref}
\usepackage{epstopdf}
\usepackage{amsmath}    
\usepackage{amssymb}    
\usepackage{xcolor}

\begin{document}
\title{Effects of strain on electronic and magnetic properties of Co/WS$_2$ junction:
       a density functional and Monte Carlo study}

\author{Hamideh Kahnouji}

\author{S. Javad Hashemifar}
\email{hashemifar@cc.iut.ac.ir}

\author{Nafiseh Rezaei}

\author{Mojtaba Alaei}

\affiliation{Department of Physics, 
Isfahan University of Technology, 84156-83111 Isfahan, Iran}

\begin{abstract}
In this work, density functional computations and Monte Carlo simulations are 
performed to investigate structural, electronic, magnetic, and thermodynamic
properties of Co/WS$_2$ junction, a semiconductor WS$_2$ monolayer covered by 
a ferromagnetic cobalt monolayer.
In addition to a conventional semilocal exchange-correlation functional,
three nonlocal functional, including the novel ACBN0 scheme,
are applied to obtain reliable electronic and magnetic properties.
It is argued that the ACBN0 scheme, is very efficient 
for first principles description of the Co/WS$_2$ junction.
The obtained electronic structures evidence a trustworthy half-metallic gap 
in the majority spin channel of the lowest energy configuration of 
the junction, promising for spintronic applications.
The obtained magnetic thermodynamics properties from Monte Carlo simulations
predict a Curie temperature of about 110\,K, which is far small
for device applications of this junction.
The electronic and magnetic properties of the system are calculated
under various compressive and tensile strains and it is shown that
a tensile strain of about 4\% may effectively improve thermal stability of
half-metallic ferromagnetism in the Co/WS$_2$ junction.

\end{abstract}
\newcommand{\etal}{{\em et al}}

\maketitle

\section{introduction}

Transition metal dichalcogenides (TMDCs) are a known class of exfolibale structures,
which are composed of sandwiched X-M-X monolayers (M=Cr,Mo,W, X=S,Se,Te).
Individual MX$_2$ monolayers, in contrast to their multilayers,
have a direct semiconducting band gap in the visible to near-infrared region 
and exhibit a high on/off switching ratio,
hence are very promising for various nanoelectronics and 
optoelectronics applications.\cite{wang2012}
Recent observation of long spin life times ($>1ns$) in a single MoS$_2$ monolayer 
has opened a promising avenue for spintronic applications of TMDCs.\cite{mak2012,xiao2012}
This long spin life time is attributed to the strong spin-valley
coupling in the monolayer, which effectively protects spin index of carriers.
As a result, spin index may act as a universal information carrier across
hybrid systems of these nanolayers with other spintronic materials.\cite{xiao2012}

However, previous studies show that ideal and pristine TMDC monolayers have
limitations for their application in spintronic devices because of 
their nonmagnetic ground state. 
In this regard, many researchers have tried to induce spin polarization
in these 2D semiconductors by using magnetic dopants, magnetic coverages, 
defects, strain, and cutting into nanoribbons.
Doping with transition metal elements is a conventional method to
obtain stable magnetism in semiconductors.
Cheng and others performed first principles computations
to identify that Mn, Fe, Co, and Zn are promising dopants
for production of 2D MoS$_2$ based diluted magnetic semiconductors.\cite{cheng2013}
Xiang \etal. experimentally confirmed room-temperature ferromagnetism in the Co doped 
MoS$_2$ nanosheets and moreover observed that increasing the doping concentration
decreases the spin polarization of the system.\cite{xiang2015}
Another first principles investigation indicates that 4-6.25\% Co substitution 
in the Mo sites of a single MoS$_2$ monolayer produces
a stable and optimized magnetic moment of 3\,$\mu_B$/unit in the system.\cite{wang2016}
Singh and colleague predicted a very rich magnetic behavior and an  
extended moment formation in a doped WS$_2$ monolayer.\cite{singh2016}

In 2014, it was theoretically predicted that a cobalt monolayer on a single MoS$_2$ monolayer
exhibits robust half-metallic ferromagnetic behavior with a complete spin-filter 
efficiency of 100\% in a broad range of bias voltage.\cite{chen2014}
Observation of half-metallic ferromagnetism over a TMDC monolayer offers
a strong evidence for promising potential of atomically thin
ferromagnet/TMDC junctions for spintronic applications.
Moreover, growth of a ferromagnetic contact on a 2D semiconductor
is likely more feasible than dilute magnetic doping of this system,
hence junctions of ferromagnetic monolayers on semiconductor TMDC monolayers 
are expected to achieve broad applications in spintronics.
Very recently, Garandel and colleagues investigated spin dependent charge transfer at 
the Co(0001)/MoS$_2$ interface by using first principles calculations
and concluded that this interface constitute an elementary block of 
a full spintronic device.\cite{garandel2017}

Recently, some theoretical studies showed that strain may also induce
and control magnetism in some 2D TMDCs.
Guo \etal. performed first principles calculations and observed that 
strain can induce or enhance spin polarization of 
TaX$_2$ (X = S, Se, Te) sheets.\cite{guo2014}
Ma and others also provided theoretical evidences for the strain tunable 
magnetization of VS$_2$ and VSe$_2$ monolayers.\cite{ma2012}
Likewise, in the hydrogenated and vacancy defected single monolayers of MoS$_2$ 
the influence of strain on the spin polarization of the system has been
theoretically justified.\cite{shi2013}
Indeed, in addition to the magnetization, engineering the band gap of 2D materials 
via strain is also one of the most viable approaches 
to achieve wide range of properties required for rational design of novel devices.

In this paper, we employ first principles computations to investigate 
possible emergence of magnetism in a single WS$_2$ monolayer covered by a 
Co monolayer and examine strain effects on electronic 
and magnetic properties of this system.

\section{Computational details}

The electronic structure calculations and structural relaxations
were performed in the framework of Kohn-Sham density functional theory (DFT)
by using the pseudopotential plane wave technique,
implemented in the {\sc Quantum-Espresso} package.\cite{quantum-espresso}
The generalized gradient functional (GGA) in the Perdew-Burke-Ernzerhof (PBE)
implementation was applied as the exchange correlation approximation,\cite{PhysRevLett.77.3865}
and the Hubbard-based DFT+U scheme was employed for better description 
of the correlated Co 3d electrons.\cite{PhysRevB.57.1505}
It should be noted that the effective Hubbard parameter $U_{eff}$ ($=U-J$)
has been used in our work, where $J$ stands for the Hund exchange.
In addition to the DFT+U method, the Heyd-Scuseria-Ernzerhof (HSE) 
hybrid functional has also been used for more accurate description 
of the electronic structure of the system.\cite{doi:10.1063/1.1564060}
The HSE calculations were conducted by using the FHI-aims package,\cite{FHI-aims}
which employs numeric atom centered orbitals for efficient full potential 
all electron description of materials and nanostructures.

A vacuum thickness of 15\,\AA\ was used to prevent artificial interaction 
by the periodic boundary conditions perpendicular to the monolayer surface.
A kinetic energy cutoff of 50\,Ry was used 
for the plane-wave expansion of the wave functions
and the Brillouin zone integrations were performed by using 
a mesh of $8\times8\times1$ $k$ points.
The total energy and atomic force convergence thresholds were set 
to $10^{-5}$\,eV and 2\,meV/\AA\ for the self-consistent calculations
and geometry optimizations, respectively.
The temperature dependence of the magnetic properties of the system
was investigated by using classical Monte Carlo (MC) simulations in 
a $20\times20\times1$ supercell containing 1600 Co atoms 
with periodic boundary conditions. 
We performed simulations with $2\times$10$^6$ MC steps for 
thermalization and $2\times$10$^6$ MC samplings for the measurement.

\section{Structural and electronic properties}

\begin{figure}
\includegraphics[scale=0.35]{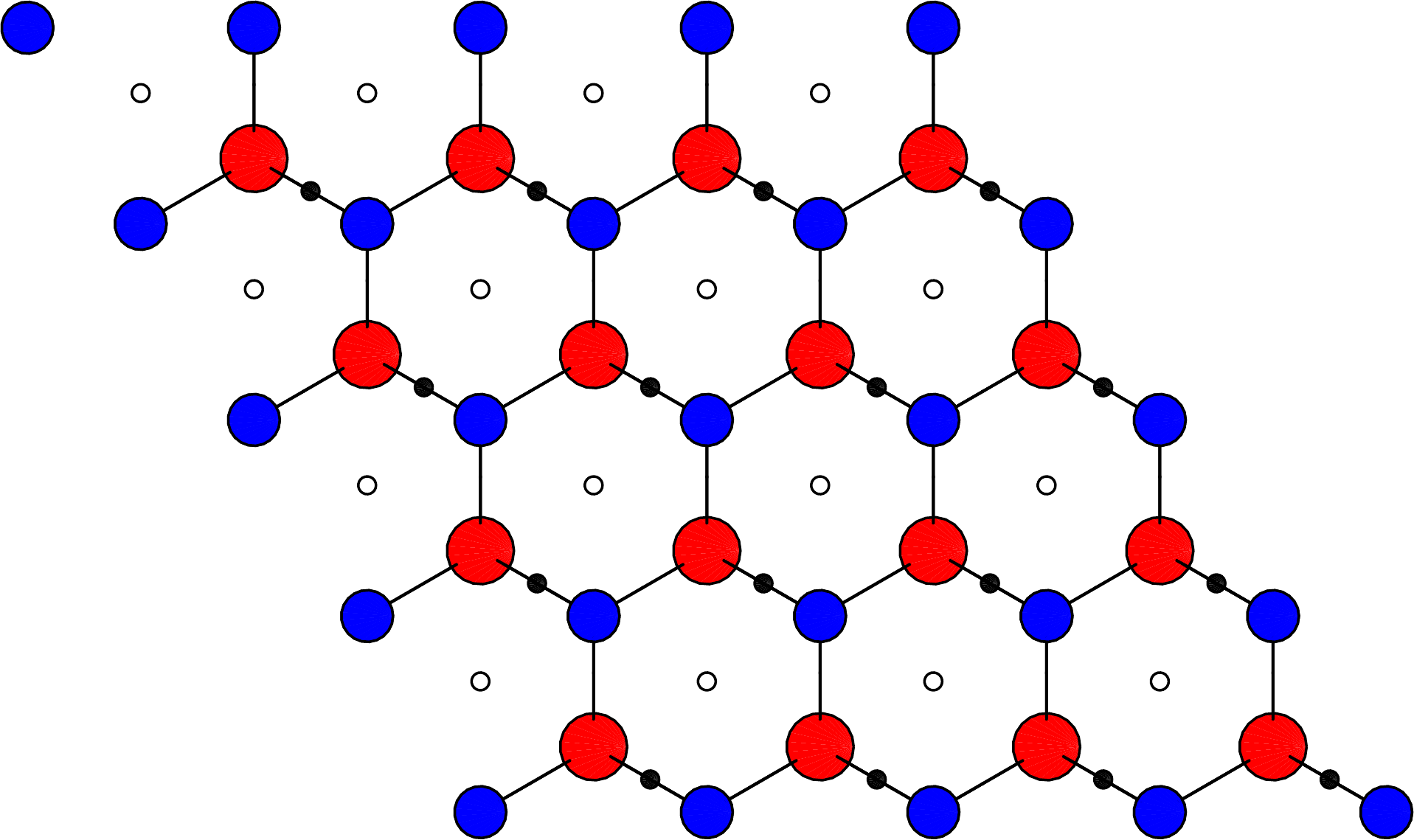} 
\caption{\label{sites}
 Top view of a WS$_2$ monolayer, 
 W and S atoms are displayed by large red and small blue balls,
 while hollow and bridge sites are shown by white and black dots,
 respectively.
}
\end{figure}

In order to study the electronic and magnetic properties of the Co/WS$_2$ junction, 
composed of a cobalt monolayer on a single monolayer of WS$_2$,
first we looked for the stable atomic structure at this interface.
In this regard, four configurations of the interface atoms were considered.
In these four configurations, which are sketched in Fig.\ref{sites},
cobalt atoms are either on W top sites (W-top), S top sites (S-top),
hollow sites (h-site), or bridge sites (b-site).
In order to determine the most favorable configuration,  
these four systems were accurately relaxed and 
their binding energy ($E_b$) were calculated as:
$$E_b = E[\rm Co/WS_2] - E[\rm Co] - E[\rm WS_2]$$
where $E[\rm Co]$ and $E[\rm WS_2]$ are the total energies 
of a free cobalt atom and a relaxed WS$_2$ monolayer, respectively. 
The calculated binding energy as well as other structural parameters of 
the investigated configurations of Co/WS$_2$ are listed in table~\ref{str}.
We observed that the Co monolayer does not stay on the bridge positions
of the WS$_2$ surface and relaxes to other configurations,
hence the b-site system is not appeared in the table.

\begin{table}
\caption{\label{str}
 Calculated binding energy $E_b$ (eV), 
 total magnetic moment per cobalt atom M ($\mu_B$), 
 interface atomic distances $d$ (\AA), S-W-S bond angle $\theta _{1} $ (degree),
 and Co-S-Co bond angle $\theta _{2} $ (degree) 
 of the three configurations of Co/WS$_2$.
}
\renewcommand{\d}[1]{$d_{\mathrm{\,#1}}$}
\begin{ruledtabular}
\begin{tabular}{lccccccc}
         & $E_b$ &  M   & \d{Co-W} & \d{Co-S} & \d{Co-Co} & $\theta_1$ & $\theta_2$ \\
\hline
 W-top   & -5.24 & 1.00 &   2.71   &   2.11   &    3.19   &     82     &   99.3 \\ 
 S-top   & -3.68 & 1.86 &   4.18   &   2.16   &    3.10   &     83     &   ---  \\
 h-site  & -3.98 & 0.00 &   3.01   &   2.05   &    3.34   &     78     &   109  \\
\end{tabular}
\end{ruledtabular}
\end{table}

Our calculations show that W-top is the most stable configuration
of the Co monolayer on WS$_2$ surface,
and the other configurations are substantially higher (1.36 and 1.64 eV) in energy,
indicating strong interface bonding in the W-top system.
In this configuration, cobalt atoms make four interface bonds,
three with the neighboring sulfur atoms and one with the underlying W atom.
The corresponding bond distances are presented in table~\ref{str}.
Moreover, we found that the strong interface bonding in the W-top system
weakens the internal bonds in the WS$_2$ monolayer and consequently
increases the W-S bond distance (2.53\,\AA) and the S-W-S bond angle (82.7$^o$),
compared to a pristine WS$_2$ monolayer (2.40\,\AA\ and 81.5$^o$). 
The integer total spin moment of 4\,$\mu_B$, obtained for the W-top configuration, 
is an evidence for the ferromagnetic half-metallic behavior of this system,
while the S-top and h-site configurations exhibit ferromagnetic
and non-magnetic metallic behaviors, respectively.

\begin{figure}
\includegraphics[trim=0 0 0 0cm,clip,scale=0.9]{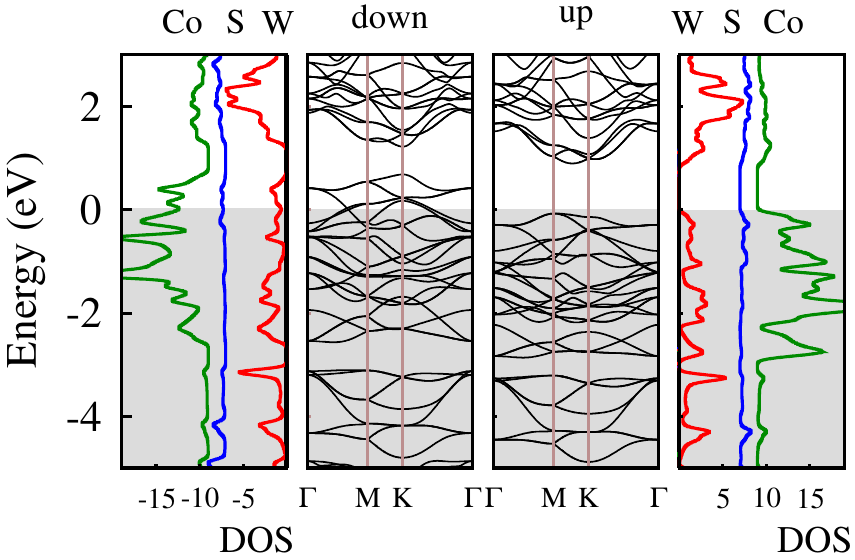}
\caption{\label{dos-pbe}
 Orbital and spin resolved density of states and band structures of 
 the stable configuration of Co/WS$_2$.
 The negative DOS belongs to the minority channel and 
 the Fermi levels are set to zero.
 The shaded areas show the occupied part of the electronic states.
}
\end{figure}

To gain further insight into the structural and magnetic behavior 
of the W-top structure of Co/WS$_2$, we investigated projected density 
of states (PDOS) and band structure of this system, illustrated in Fig.~\ref{dos-pbe}. 
It reveals that the system posses a half-metallic electronic structure
which is very promising for spintronic applications.
The half-metallic gap appears in the majority channel
while the 100\% spin polarized charge carriers are fully 
accommodated in the minority channel.
It is seen that the primary contributions to the valence states
arise from the Co 3d orbitals, with an exchange splitting of about 1\,eV. 
The results show that Co atoms introduce partially occupied bands into 
the gap of WS$_2$. In the minority spin channel, the 3p states of S atom 
and the 5d states of W atom contributes significantly to the unoccupied states 
and indicate a strong hybridization between Co and its neighboring S and W atoms.
Furthermore, the strong hybridization of the Co atom with its neighboring S atoms, 
leads to a ferromagnetic (FM) superexchange interaction in the Co/WS$_2$ junction
(will be explained in the next section).

In order to confirm the ferromagnetic ground state of the favorable 
configuration of Co/WS$_2$, the total energy of this system 
was minimized by starting from an antiferromagnetic (AF) state,
constructed in a $2\times2\times1$ lateral supercell.
We observed that this calculation converges to a nearly non-magnetic 
state with about 77\,meV higher energy, compared with the FM state,
indicating favorability of half-metallic ferromagnetism
in the stable configuration (W-top) of a Co monolayer on a WS$_2$ monolayer.
One may also expect non-trivial influence of spin-orbit coupling (SOC)
on the stable spin configuration of the system.
Therefore, we carried out DFT+SOC calculations and obtained 
a metastable AF state for the W-top configuration,
which is about 70\,meV less stable than the FM state.
Hence, ferromagnetic ordering is the true magnetic ground state
of the stable configuration of Co/WS$_2$,
although, spin-orbit coupling enhances feasibility
of formation of AF spin ordering in the system. 
Experimental observations confirm occurrence of ferromagnetism
in the similar Co/WSe$_2$ and Co/Mo$S_2$ junctions.\cite{ahmed2018,xiang2015}
The observed FM ground state of the Co/WS$_2$ junction
is likely due to the superexchange interaction between neighboring 
cobalt atoms in the Co monolayer,
because of the rather large distance ($\sim3.18$\,\AA) between these Co atoms.
On the other hand, the close Co-S distance of about 2.11\,\AA\
shows that S atoms are able to mediate a superexchange interaction
between cobalt atoms.
According to the Kanamori-Goodenough rule,\cite{owen1966,goodenough1955}
if the Co-S-Co bond angle is around 90$^{\circ}$, 
a ferromagnetic superexchange interaction is expected between Co atoms,
in well agreement with our calculated bond angle (table~\ref{str}).
If this bond angle is in the range of 120-180$^{\circ}$,
the Kanamori-Goodenough rule predicts an antiferromagnetic superexchange 
interaction in the system.

\section{Effects of nonlocal correlations}

Theoretical study of the Co/WS$_2$ junction is complicated by the correlated 
nature of the cobalt 3d electrons.
Conventional local and semilocal DFT functionals (LDA/GGA),
are known to suffer from the self-interaction error,
which enhances delocalization of the electronic wave functions
and hence prevents a satisfactory description of 
the localized d or f states.
To overcome this problem different corrections to the PBE wavefunctions are adopted.
The DFT+U scheme is a popular method for better description 
of these localized orbitals, although proper identification of
the Hubbard U parameter is an elaborating task.
In older schemes, the Hubbard parameter is scanned in a proper range to recover 
some relevant experimental data, while in modern schemes, 
first principles determination of this parameter is more appreciated.
Because of the lack of proper experimental data on Co/WS$_2$,
we focused on the first principles determination of U 
in the framework of linear response approach.\cite{PhysRevB.71.035105}
The required calculations was performed for the Co 3d states in 
a $4\times4\times1$ supercell with 64 atoms.
This method gives a self-consistent value of 5.9\,eV for the effective U parameter.

\begin{figure*}
\begin{tabular}{ccc}
   DFT+U  &  ACBN0  & HSE  \\
\includegraphics[scale=0.70]{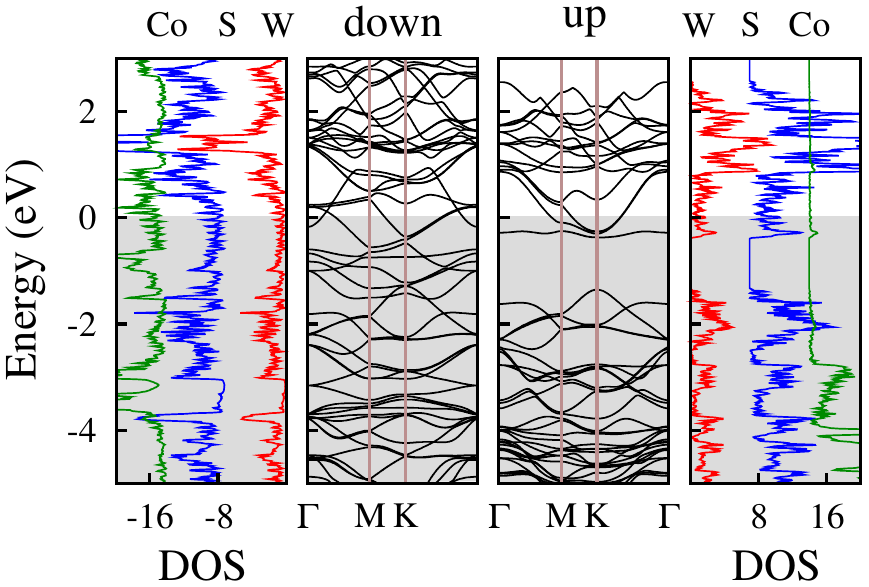} &
\includegraphics[scale=0.70]{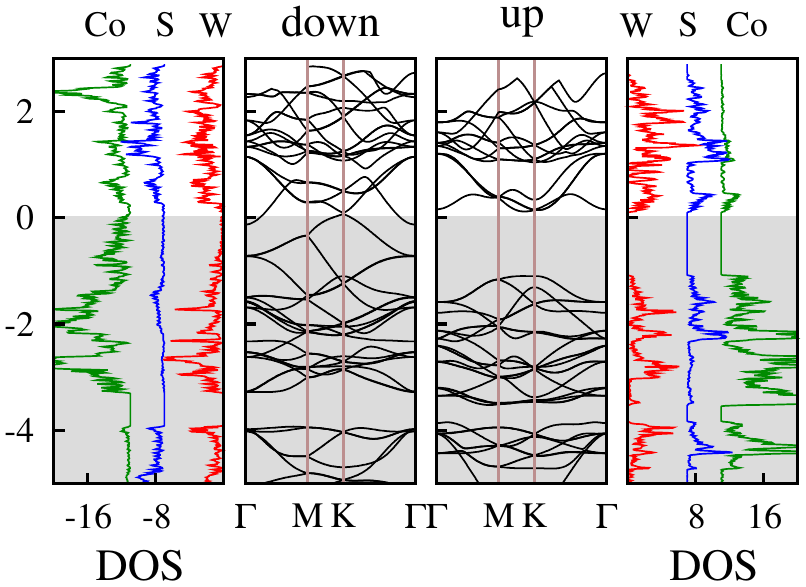}& 
\includegraphics[scale=0.70]{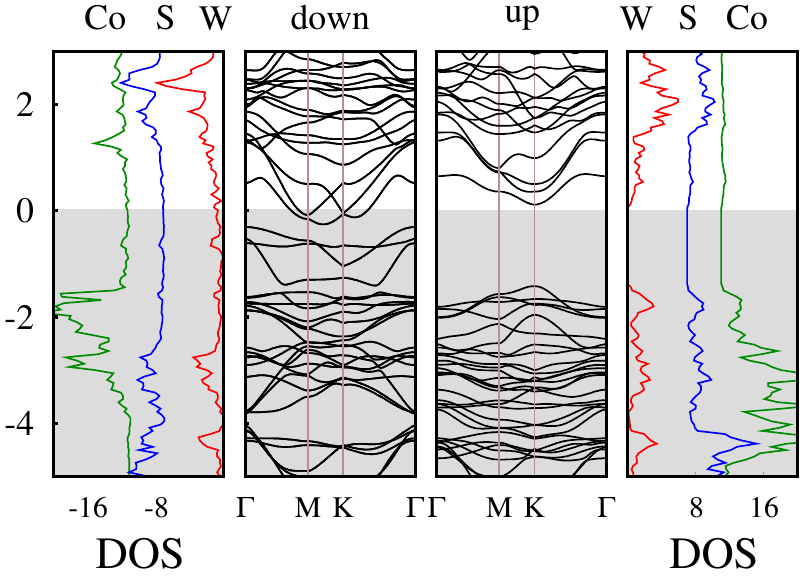} \\
\end{tabular} 
\caption{\label{dos-u}
  Obtained spin resolved electronic band structure and 
  corresponding density of states (DOS) of Co/WS$_2$ 
  within the DFT+U, ACBN0, and HSE methods.
  The spin up and down DOS plots are resolved into atomic contributions.
  The Fermi levels are set to zero and the gray regions
  show the occupied part of the electronic states.
}
\end{figure*}

The ACBN0 functional, recently proposed by Curtarolo and Nardelli,\cite{PhysRevX.5.011006}
is a novel pseudo-hybrid Hubbard density functional for efficient and accurate 
DFT+U calculations with no need to any empirical parameter.
For a series of transition metal oxides and chalcogenides, 
it was shown that the obtained structural, electronic, and magnetic properties 
within ACBN0, compare well with the first principles GW calculations and 
experimental measurements, at a fraction of the computational cost.\cite{acbn0-2015,acbn0-2017}
We used the computer package {\sc AFLOW$\pi$} for our DFT+U calculations
in the ACBN0 scheme.\cite{supka2017}
The obtained converged Hubbard values in this scheme are
U$_{\rm Co-3d}=4.65$, U$_{\rm S-3p}=1.24$, and U$_{\rm W-5d}=0.17$\,eV. 
Contrary to the predominant focus on the partially Co 3d states, 
the ACBN0 results suggest a small Hubbard correction on the sulfur 3p 
as well as on the close shell W 5d states.
Some other theoretical studies have also argued the importance of 
the Hubbard correction to the p states for correct description of 
the electronic structure of TiO$_{2}$, ZnO, CdO  and other compounds.
\cite{mattioli2014,acbn0-2015,cao2008}

The computationally expensive method of hybrid functionals is an alternative approach 
for better description of the electronic structure of strongly correlated compounds. 
These functionals take into account nonlocal electronic effects by mixing 
a fraction of the Hartree-Fock exchange with the local or semilocal DFT functionals.
In the current project, we have performed certain time consuming computations within 
the HSE06 hybrid functional to compare with our ACBN0 and DFT+U calculations.

The obtained electronic structure of Co/WS$_2$ within the DFT+U, ACBN0, 
and HSE methods are presented in Fig.~\ref{dos-u}.
It is seen that DFT+U predicts a full metallic state for the junction,
while ACBN0 and HSE detect a nontrivial half-metallic gap in the system.
In the absence of proper experimental data, 
we attribute the metallic behavior of Co/WS$_2$ within DFT+U
to the overestimated value of the U parameter (5.9\,eV) in this scheme.
One may find some other cases in the literature where the linear response approach
has overestimated the Hubbard parameter of the system.\cite{Linscott-2018}

It is seen that the ACBN0 and HSE functionals, 
similar to the PBE results presented in Fig.~\ref{dos-pbe},
predict a half-metallic behavior for the Co/WS$_2$ junction.
The obtained half-metallic band gaps, presented in table~\ref{methods},
show that HSE gives the highest half metallic gap while PBE leads the smallest one. 
Some theoretical studies argue that the HSE functional 
may systematically overestimate the band gap,\cite{kang2013,li2013,wang2013,lu2014mos} 
while the semilocal PBE functional usually underestimates this parameter.
Hence one may expect that ACBN0 gives a more reliable half-metallic 
band gap for Co/WS$_2$.
The obtained electronic structure, presented in Fig.\ref{dos-u},
indicate good similarity between the general features of density of states 
within HSE and ACBN0, while corresponding electronic band structures exhibit 
some differences around the Fermi level.

In order to calculate magnetic thermodynamic properties of the system,
one has to perform extensive statistical averaging over a large ensemble of 
spin configurations of the system, by using the Monte Carlo scheme.
To facilitate this procedure, the energy of individual spin configurations
is usually parameterized in terms of magnetic exchange parameters $J$ via
the Heisenberg Hamiltonian $H=-\sum J_{ij}S_i\cdot S_j$.
The exchange parameters are conventionally calculated from the vertical 
energy difference of various spin configurations of the system
with the ferromagnetic state.
It means that the relaxed configuration of the system in the FM state
is used to construct proper supercells for total energy calculation 
of various spin configurations.
We used a 2$\times$2 lateral supercell for calculation of exchange parameters.
It should be noted that only one antiferromagnetic (AF) spin arrangement of cobalt atoms 
is applicable in a 2$\times$2 supercell,
hence, we only considered the nearest-neighbor Co-Co interaction 
for our thermodynamics study.
The influence of the next nearest neighbor couplings will be estimated.
Comparing the energy difference of the AF and FM states
gives the nearest neighbor Co-Co exchange parameters,
presented in table~\ref{methods}.
Among the four investigated methods, the best agreement is seen
between HSE and ACBN0.
It is seen that the nonlocal effects presented in DFT+U, HSE, and ACBN0,
compared with PBE, weakens the exchange interaction between neighboring cobalt atoms.
It is likely attributed to the attenuated hopping amplitude of the neighboring d electrons 
at the expense of enlarged on-site Coulomb repulsion in the nonlocal functionals.
In other words, the more localized nature of the cobalt 3d electrons 
within DFT+U, HSE, and ACBN0, reduces the corresponding exchange overlap integrals,
compared with the semilocal PBE functional.

\begin{figure}
\includegraphics[scale=0.9]{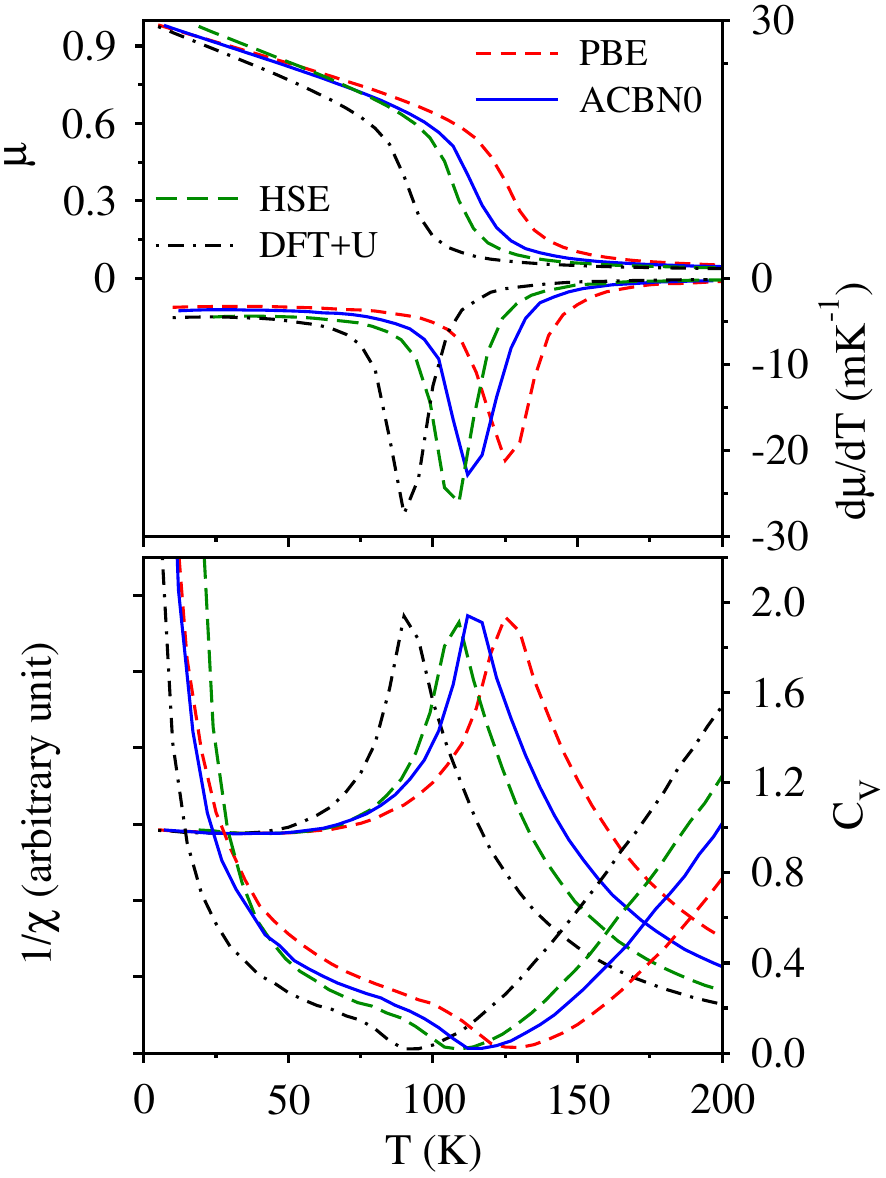} 
\caption{\label{cv}
  Obtained magnetic order parameter $\mu$ and its temperature derivative,
  specific heat $\rm C_V$, and susceptibility inverse ($1/\chi$) of the Co/WS$_2$ junction
  within the PBE, DFT+U, ACBN0, and HSE methods.
}
\end{figure}

The obtained exchange parameter was used for classical MC calculation
of the thermodynamic parameters heat capacity $C_v$, susceptibility inverse $1/\kappa$,
and average magnetic moment $\mu$ in the temperature range of 5 to 240\,K,
with temperature steps of 5\,K.
The obtained results are presented in Fig.~\ref{cv}.
The calculated temperature derivative of the average magnetic moment as well as 
the heat capacity clearly show the position of the magnetic phase transition of the junction.
The extracted Curie transitions temperatures, listed in table~\ref{methods},
are about 110\,K which is not large enough for anticipated 
device applications of this junction.
In the next section, we will show that a uniform strain of about 4\%
may be utilized to enhance the Curie temperature toward room temperature.
Once again we observe that among the four investigated methods,
HSE and ACBN0 give the closest transition temperatures.

\begin{table}
\caption{\label{methods}
 Obtained nearest neighbor exchange parameter J (meV), 
 Curie temperature T$\rm_C$ (K),
 half metallic gap (eV), and equilibrium lattice parameter $a$ (\AA)
 within four investigated schemes.
}
\begin{ruledtabular}
\begin{tabular}{lcccc}
       & $J$ & T$\rm_C$ &  gap  & $a$  \\
\hline                           
 PBE   & 8.3 &   126    & 1.01  & 6.36 \\
 ACBN0 & 7.3 &   115    & 1.28  & 6.39 \\
 HSE   & 6.9 &   109    & 1.45  & 6.41 \\
 DFT+U & 6.0 &    95    &  ---  & 6.47 \\
\end{tabular}
\end{ruledtabular}
\end{table}

As it was mentioned, we have only considered the nearest neighbor exchange
interaction for our thermodynamic study, because calculating next nearest 
neighbor couplings requires larger supercells.
In order to estimate the error of this procedure,
two different AF spin configurations were considered in a 4$\times$4 supercell,
within the PBE scheme.
The obtained total energies are compared with that of the FM state 
to achieve nearest neighbor and next nearest neighbor exchange parameters 
$J_1=6.3$ and $J_2=2.3$\,meV, respectively.
After preforming proper MC calculations, the new Curie temperature
of the system within PBE was found to be about 140\,K,
only 14\,K higher than the result obtained by considering
only the nearest neighbor exchange interaction (table~\ref{methods}).
It is already discussed that changing the number of Heisenberg exchange neighbors,
although influence individual exchange parameters,
has weak effect on the resulting thermodynamic properties.\cite{askari}
Hence, in this work, in order to prevent heavy supercell calculations,
we limit our thermodynamics study to the nearest neighbor exchange interaction and
assign an inaccuracy of about 10-20\,K to the obtained transition temperatures.

The overall conclusion of this section is that ACBN0 is an efficient scheme for 
first principles calculation of electronic and magnetic properties of the Co/WS$_2$ junction. 
The main reason is the close agreement observed between ACBN0 and the expensive HSE functional 
in determination of density of states, exchange parameter, and Curie temperature.
Moreover, it was argued that the obtained half-metallic gap within ACBN0 is likely
more accurate, compared with HSE and PBE results.
Hence, in the next section, we will concentrate on the ACBN0 method
for the study of strained Co/WS$_2$ junction.
It was also shown that, within the PBE scheme, 
the spin-orbit coupling may influence the energy difference 
between FM and AF states of the junction, hence, 
magnetic exchange parameters and consequently magnetic thermodynamics
properties may be sensitive to this relativistic correction.
In this regard, we considered the spin orbit coupling in the ACBN0 scheme
and calculated the energy difference of the FM and AF states by ACBN0+SOC method.
The obtained exchange coupling constant (6.8\,meV) show slight effect of 
the SOC correction in the magnetic thermodynamics properties of Co/WS$_2$.

\section{Effects of strain}

\begin{figure}
\includegraphics[scale=0.9]{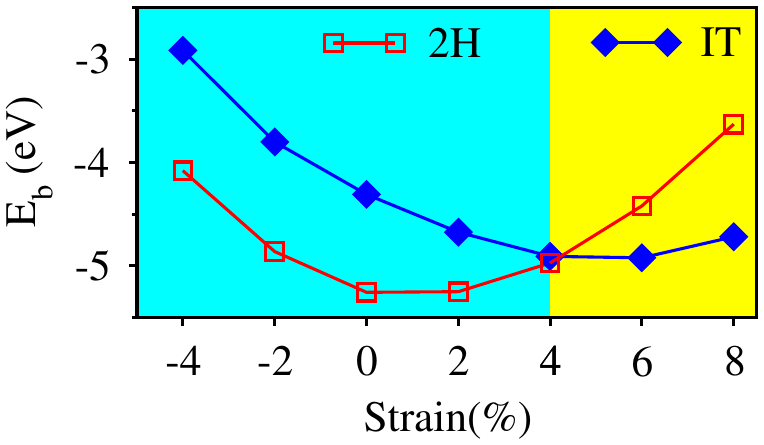}
\caption{\label{1T2H}
 Effect of biaxial strain on the binding energy of 
 the H and T phases of Co/WS$_2$.
}
\end{figure}

In the previous section, it was seen that the predicted Curie temperature
of Co/WS$_2$ junction ($\sim110$\,K) is very small for most technological applications.
Hence, in this section, we investigate whether strain may be helpful
to improve magnetic properties of this junction toward practical purposes.
The strained junctions are calculated within the ACBN0 scheme,
although the PBE results are also provided for comparison.
The magnitude of strain $\varepsilon$ is defined as
$\varepsilon =[(a - a_{0})/a_{0}]$ 
where $a$ and $a_0$ are the lattice parameters of the strained and 
equilibrium structures, respectively. 
The atomic positions within all strained structures were fully relaxed.
The first issue to be considered, is the stability of the junction under strain.
Individual WS$_2$ sheets, in general, exhibit two polymorphs:
trigonal prismatic (H-phase) and octahedral coordinated (T-phase).
The most stable crystal structure is the semiconductor H phase,
while the metastable T phase exhibits a metallic behavior.\cite{T-phase} 
First, we investigated whether a biaxial strain may induce 
a structural phase transition from the stable H-phase to the metastable T-phase
in a WS$_2$ monolayer covered by a cobalt monolayer.
In this regard, binding energy of the junction in both phases
was calculated under various biaxial tensile and compressive strains.
The results, shown in Fig.~\ref{1T2H},
indicate that a biaxial tensile strain of more than 4\%  
may convert the WS$_2$ structure of the junction to the T-phase. 
Therefore, in the following, we study the strain values between -4\% to 4\%
to prevent the possible structural phase transition in the system.

\begin{figure}
\includegraphics*[scale=0.85]{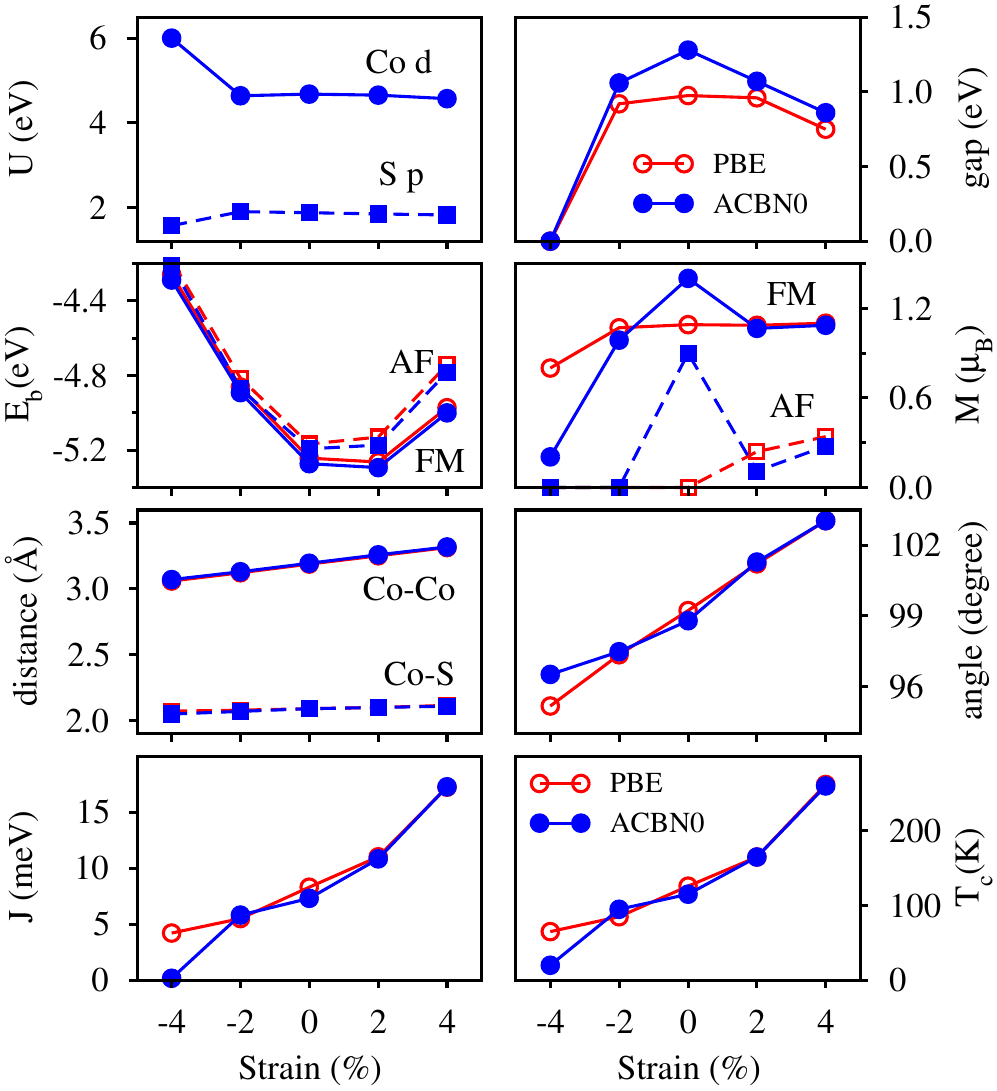}
\caption{\label{strain}  
 Obtained effective Hubbard U parameters in the ACBN0 scheme,
 half-metallic band gap, binding energy $E_b$ of the FM and AF states,
 absolute magnetic moment of the system $M$ per cobalt atoms,
 Co-Co and Co-S bond distances, Co-S-Co bond angle,
 magnetic exchange parameter J,
 and Curie temperature (T$_c$) of Co/WS$_2$
 as a function of biaxial strain.
} 
\end{figure}

Various calculated physical properties of the strained Co/WS$_2$ junction
are presented in Fig.~\ref{strain}. 
We observe that the obtained effective U parameters are almost insensitive 
to the strain for the strain values larger than -2\%,
while at the strain of -4\% a rather large U parameter is obtained
for the cobalt d states,
indicating the highest localized nature of the Co d states in this system.
This high localization quenches the magnetic moment of cobalt atoms and
close the half-metallic gap of the -4\% strained junction (Fig.~\ref{strain}).
Because within the ACBN0 scheme, the half-metallic gap of the junction is
enclosed by the cobalt d states (Fig.\ref{dos-u}) and the magnetic moment
of the cobalt atoms is also originated from d electrons.
In order to inspect the effect of strain on the magnetic state of the system,
we investigated the strained Co/WS$_2$ junction within 
both ferromagnetic and antiferromagnetic states. 
The obtained binding energies, presented in Fig.\ref{strain},
indicate that tensile strain increases the relative stability of the FM state,
while compressive strain reduces the energy distance between FM and AF states.
The calculated magnetic moments show that compressive strain pushes
both FM and AF spin orderings toward a nonmagnetic state.
It is due to the fact that compressive strain enhances crystal field of 
the system and hence weakens the spin polarization of the cobalt layer.

As shown in the figure, within both ACBN0 and PBE, 
the Co-Co distance is more sensitive to the strain, compared with the Co-S distance, 
because Co atoms are in the same plane while S atoms are in a separate layer 
close to the Co layer.
Therefore, tensile strain is expected to enhance the role of FM superexchange 
interaction between cobalt atoms in the magnetic behavior of the junction.
The faster increase of the Co-Co distance, compared with the Co-S distance,
leads to increasing the Co-S-Co bond angle by increasing strain,
illustrated in Fig~\ref{strain}.
However, obtained bond angle of the strained junctions remains well 
in the ferromagnetic region of the Kanamori-Goodenough rule.
The half metallic gap of the system is found to decrease 
by both tensile and compressive strains,
in such a way that the equilibrium structure gives the highest
half metallic gap within ACBN0 and PBE.
The observed decrease under compressive strain is faster
and the gap is closed at the strain of -4\%,
while tensile strain up to 4\% preserve the half metallic nature of the junction.
Compared with PBE, ACBN0 gives a larger half metallic gap, 
which is expected to be more close to the reality.

\begin{figure}
\includegraphics[scale=0.8]{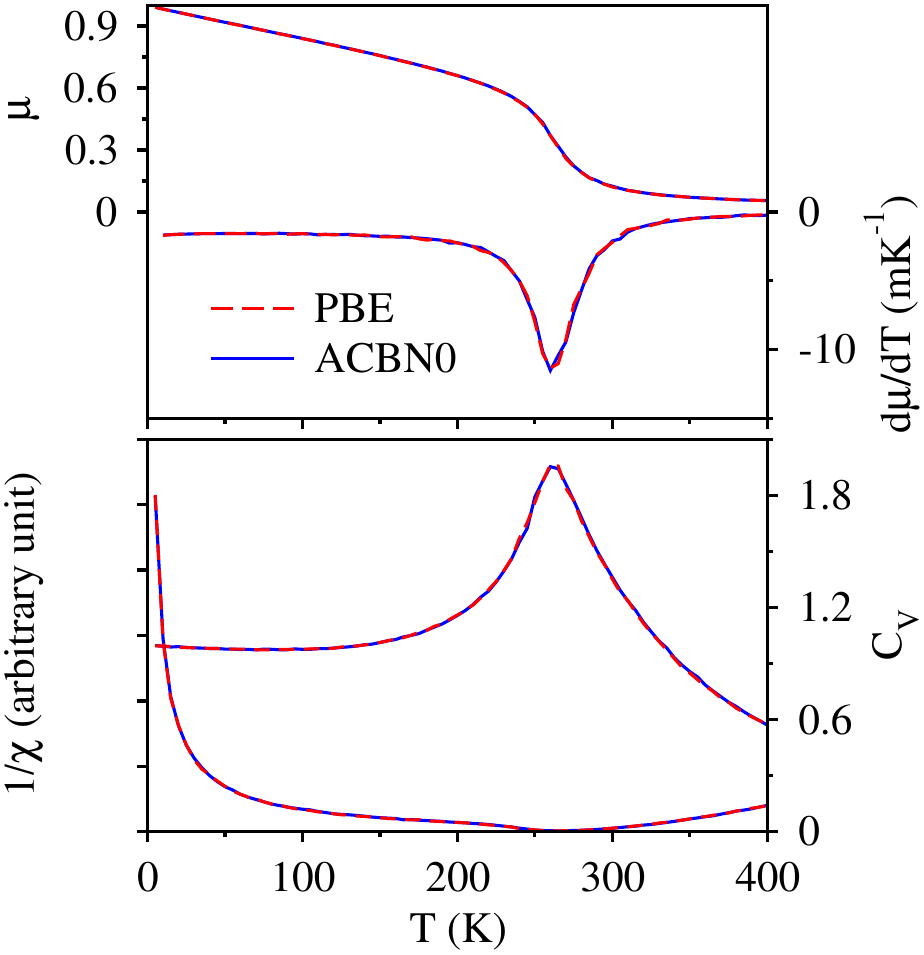} 
\caption{\label{cv4}
  Obtained magnetic thermodynamic parameters of the Co/WS$_2$ junction
  under 4\% tensile strain within PBE and ACBN0.
}
\end{figure}

At the final step, we consider the effects of strain on magnetic thermodynamic
properties of the stable configuration of Co/WS$_2$ junction.
In this regard, first, the magnetic exchange parameters
were calculated by computing the vertical energy difference between
the FM and AF states of the strained systems.
Then, with the same setup as equilibrium structure,
classical MC calculations were performed to obtain
magnetic thermodynamic properties of the strained junctions.
The obtained magnetic exchange parameters and Curie temperatures,
as a function of strain, are given in Fig.~\ref{strain}.
We can see that within both PBE and ACBN0 functionals, 
exchange coupling $J$ and consequently the Curie temperature of 
the Co/WS$_2$ junction increase with increasing strain.
The Curie temperature under a tensile strain of 4\% reaches 
a close room temperature value of about 262\,K.
This indicates that tensile strain is promising
for realization of the potential device applications of the Co/WS$_2$ junction.
The same enhancement is experimentally observed
in the Curie temperature of the strained FeNi thin films.\cite{fe64ni36}

The magnetic thermodynamic properties of the 4\% strained junction,
as a function of temperature, are given in Fig.~\ref{cv4}.
We observe that the PBE and ACBN0 results coincide very well
at this value of strain. 
Almost all obtained properties, presented in Fig.~\ref{strain},
confirm that by increasing strain the difference between 
the PBE and ACBN0 descriptions decreases.
It is more visible in the trend of Co-S-Co bond length,
exchange coupling parameter, and Curie temperature.
In a recent preprint,\cite{kim2018} 
it is discussed that weakening of the Hubbard interaction 
under strain is an screening effect.
In other words, tensile strain increases p-d band overlap
and consequently enhances screening effects in the system.
\newline\newline

\section{Conclusions}

In summary, we employed pseudopotential DFT calculations to investigate 
structural, electronic, and magnetic properties of a Co monolayer on 
a semiconductor WS$_2$ monolayer.
It was seen that cobalt atoms prefer to stay on top of W atoms 
to make a triangular pyramid with three neighboring surface S atoms.
The electronic and magnetic properties of the stable configuration of
the junction were considered within the semilocal PBE and nonlocal
DFT+U, ACBN0, and hybrid HSE schemes.
The obtained electronic structures indicate that PBE, ACBN0, and HSE
predict a ferromagnetic half-metallic ground state for the Co/WS$_2$ junction,
while DFT+U gives a metallic state, which is due to 
the overestimated value of the effective Hubbard parameter in the linear response approach.
Next, by calculating the vertical energy distance between the ferromagnetic
and antiferromagnetic states, the nearest neighbor exchange coupling between 
cobalt atoms were extracted and used for Monte Carlo simulation of 
the thermodynamics properties of the junction.
The obtained temperature profile of the magnetic order parameter and 
heat capacity indicate a transition temperature of about 110\,K for the system.
Comparing the obtained electronic and magnetic properties within PBE, DFT+U,
ACBN0, and HSE indicates that the novel ACBN0 scheme is a very efficient
method for first principles simulation of the Co/WS$_2$ junction.
Hence, this scheme was used to investigate the effect of tensile and compressive
strains on the electronic and magnetic properties of the system.
It was shown that a tensile strain of about 4\%, may effectively
enhance the transition temperature of the junction toward room temperature,
while preserving the half-metallic behavior of the system.

\end{document}